\documentclass[english]{llncs}
\usepackage[T1]{fontenc}
\usepackage[latin9]{inputenc}
\usepackage{units}
\usepackage{textcomp}
\usepackage{mathrsfs}
\usepackage{amsmath}
\usepackage{amssymb}

\makeatletter

\usepackage{graphicx}
\usepackage{xcolor}
\definecolor{darkblue}{rgb}{0,0,0.5} 

\usepackage[margin=1in]{geometry}

\makeatother

\usepackage{babel}

\begin{document}

\title{Heuristic Rating Estimation Approach to The Pairwise Comparisons
Method}

\author{Konrad Ku\l{}akowski }

\institute{Department of Applied Computer Science,\\
AGH University of Science and Technology\\
Al. Mickiewicza 30, \\
30-059 Cracow, Poland\\
\email{konrad.kulakowski@agh.edu.pl}}

\maketitle

\begin{abstract}
The Heuristic Ratio Estimation\emph{ (HRE)} approach proposes a new
way of using the pairwise comparisons matrix. It allows the assumption
that the weights of some alternatives (herein referred to as concepts)
are known and fixed, hence the weight vector needs to be estimated
only for the other unknown values. The main purpose of this paper
is to extend the previously proposed iterative \emph{HRE} algorithm
and present all the heuristics that create a generalized approach.
Theoretical considerations are accompanied by a few numerical examples
demonstrating how the selected heuristics can be used in practice. 
\end{abstract}

\section{Introduction\label{sec:Introduction}}

The first evidence of the usage of pairwise comparisons (herein abbreviated
as\emph{ }PC) comes from \emph{Ramon Llull} (the XIII century) \cite{Colomer2011rlfa,Schlager2001sait2},
then the method was rediscovered in the XIX century by \emph{Fechner}
\cite{Fechner1966eop}. In the first half of the twentieth century
it was developed by \emph{Thurstone} \cite{Thurstone27aloc}. The
Analytic Hierarchy Process \emph{(AHP)}, introduced by Saaty \cite{Saaty2008rmai},
was another important extension to the PC theory, providing handy
methods for dealing with the large number of criteria.  Many examples
demonstrate the usefulness of the method \cite{Vaidya2006ahpa,Ho2008iahp,Liberatore2008tahp,Subramanian2012aroa}.
Despite its long existence, research in the field of the PC research
is still conducted. This is evidenced by the works discussing the
strengths and weaknesses of the most popular \emph{AHP} approach \cite{Dyer1990rota,Saaty1990aeot,Barzilai1994arrn,Saaty1998rbev,BanaeCosta2008acao},
and also by the works proposing the new \emph{PC} paradigms, and exploring
the new areas of applicability, such as the \emph{Rough Set} theory
approach \cite{Greco2011fk}, fuzzy \emph{PC} relation handling \cite{Mikhailov2003dpff,Fedrizzi2010otpv,Yuen2012mmpm,Yuen2013fcnp},
incomplete \emph{PC} relation \cite{Bozoki2010ooco,Fedrizzi2007ipca,Koczkodaj1999mnei},
data inconsistency reduction \cite{Koczkodaj2010odbi}, non-numerical
rankings \cite{Janicki2012oapc} and others. A broader discussion
of the \emph{PC} method can be found in \cite{Smith2004aada,Ishizaka2009ahpa}. 

The newly proposed \emph{HRE} approach \cite{Kulakowski2013ahre}
explores the use of the \emph{PC} method in cases when some alternatives
(herein referred to as concepts) have known and fixed priorities.
Therefore, it divides the concepts into two sets - initially known
elements $C_{K}$ for which the weights are fixed and unknown elements
$C_{U}$ for which the weights need to be determined. Then, by iteratively
averaging the available weights (initially only the weights of elements
from $C_{K}$ are available), subsequent propositions of the weight
vector are computed. 

The notion inherently integrated with the \emph{PC} method is data
inconsistency \cite{Brunelli2013iifp,Bozoki2008osak}. If the data
are fully consistent then any single comparison provides enough information
about the relative order and the intensity of preferences of two concepts.
In such a~case, after performing $n-1$ comparisons, the weights
of all $n$ concepts can be easily determined, provided that the n
\textminus{} 1 comparisons involve all the concepts. Thus, any special
way of data processing in order to derive the priorities of concepts
is not needed. If the input data are inconsistent the best thing to
be done is to propose a heuristic that, despite the data inconsistency,
allows the weights of concepts to be calculated.

The presented work is a follow-up of \cite{Kulakowski2013ahre}. It
introduces the new ideas (the lack of reciprocity or the lack of data)
and presents the concepts introduced in \cite{Kulakowski2013ahre}
in a more systematic and formal way. The \emph{HRE} approach presented
in this article includes four complementary heuristics  that are
useful for calculating weights when the reference set of initially
known elements $C_{K}$ is given (Sec. \ref{sec:Original-HRE-Algorithm}).
Besides theoretical consideration the article examines the \emph{HRE}
weight derivation procedures on a few numerical examples (Sec. \ref{sec:Numerical-examples}).
The article is opened by two sections introducing the \emph{PC} method
(Sec. \ref{sec:Introduction} and \ref{sec:A-pairwise-comparisons}).
A brief summary is provided in (Sec. \ref{sec:Summary}). Additional
explanations and definitions are placed in the appendices.

\section{A pairwise comparisons method\label{sec:A-pairwise-comparisons} }

Man always has to make choices. Therefore he/she always has to make
comparisons. The best bet is when one (the better one) needs to be
selected from a pair. People are accustomed to this type of comparison.
In daily contact, in the market, where paying for a fruit everyone
is trying to choose the heavier one. The relative weight of two fruits
that look like they are the greatest can be easily estimated by comparing
the weight of the fruit held in one hand with the weight of the fruit
held in the other hand. Usually, making the right choice is possible
without any additional tools indicating weight. In reality people
have to compare much more complicated things than fruits. Often there
is no way to make an accurate comparison. There is no 'weight' for
the problem. Even worse, usually there are many different things that
need to be compared. In such a~case the PC approach comes to the
rescue. It allows people to do what they do best - comparing pairs.
The final synthesis of partial assessments is performed in accordance
with predefined algorithms, such as the eigenvalue method or geometric
mean method \cite{Ishizaka2006htdp}. 

\begin{figure}
\begin{centering}
\centering 
\def\svgwidth{0.80\columnwidth} 
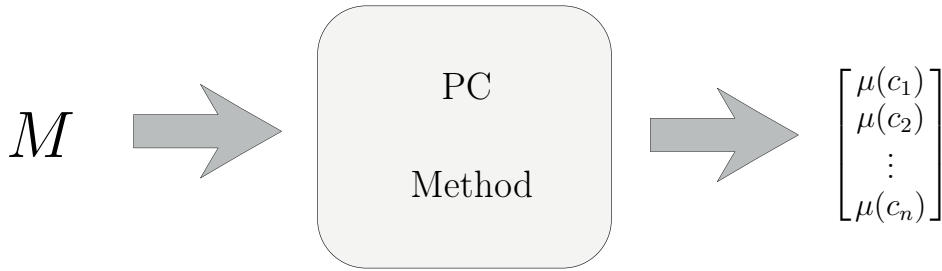
\par\end{centering}

\caption{PC Method input-output scheme}
\label{fig:PCMethodScheme}

\end{figure}

The input data to the \emph{PC} method is a \emph{(PC)} matrix $M=(m_{ij})$
and $m_{i,j}\in\mathbb{R}_{+}$ where $i,j\in\{1,\ldots,n\}$ represents
partial assessments over the finite set of concepts $C\overset{df}{=}\{c_{i}\in\mathscr{C},\, i\in\{1,\ldots,n\}\}$
where $\mathscr{C}\neq\emptyset$ is a universe of concepts. Let $\mu:C\nrightarrow\mathbb{R}_{+}$
be a partial function that assigns to some concepts from $C\subset\mathscr{C}$
positive values from $\mathbb{R}_{+}$. Thus, the value $\mu(c)$
represents the importance of $c$. The output of the \emph{PC} method
is the function $\mu$ defined for all $c\in C$. It introduces the
total order in $C$ and usually will be written in the form of a vector
of weights $\mu\overset{df}{=}\left[\mu(c_{1}),\ldots\mu(c_{n})\right]^{T}$
(see Fig. \ref{fig:PCMethodScheme}). 

Concepts, originally referred to in the literature as subjective stimuli
\cite{Thurstone27aloc}, alternatives \cite{BanaeCosta2008acao} or
activities \cite{Saaty1977asmf}, represent objects for which the
relative importance indicators $m_{ij}$ and $m_{ji}$ need to be
assessed.

It is assumed that, according to the best knowledge of experts, the
importance of $c_{i}$ equals $m_{ij}$ of the importance of $c_{j}$
i.e. $\mu(c_{i})=m_{ij}\mu(c_{j})$. The matrix $M$ is said to be
reciprocal if $\forall i,j\in\{1,\ldots,n\}:m_{ij}=\frac{1}{m_{ji}}$.
This property reflects the intuition that if the relative importance
ratio $c_{i}$ to $c_{j}$ is $m_{ij}$ then the the importance ratio
$c_{j}$ to $c_{i}$ should be $\nicefrac{1}{m_{ij}=m_{ji}}$. However,
intuitive reciprocity may not always be met. The matrix $M$ without
reciprocity property is sometimes referred to in the literature as
a~generalized \emph{PC} matrix \cite{Koczkodaj1999caca}.

Ideally $M$ is also consistent i.e. $\forall i,j,k\in\{1,\ldots,n\}:m_{ij}\cdot m_{jk}\cdot m_{ki}=1$.
Unfortunately, the knowledge stored in the \emph{PC} matrix usually
comes from different experts, the consistency condition may not be
met. In such a case, reasoning using $M$ may give ambiguous results.
This leads to the data consistency (and inconsistency) concept formalized
in the form of the inconsistency index. There are several different
inconsistency indexes, including the \emph{Eigenvector Method} \cite{Saaty2008rmai},
\emph{Least Squares Method}, \emph{Chi Squares Method} \cite{Bozoki2008osak},
\emph{Koczkodaj's} \emph{distance based inconsistency index }\cite{Koczkodaj1993ando}
and others. The most popular eigenvalue based approach \cite{Saaty2008rmai}
defines the consistency index (sometimes referred as the consistency
ratio) as 
\begin{equation}
CI=\frac{\lambda_{\text{\textit{max}}}-n}{n-1}\label{eq:Consistency_Index_AHP}
\end{equation}

where $\lambda_{\text{max}}$ is the principal eigenvalue of $n\times n$
matrix $M$. The iterative \emph{HRE} algorithm \cite{Kulakowski2013ahre}
adopts the last of them as a convenient and easy to use 'gauge' of
data inconsistency. Koczkodaj's inconsistency index $\mathscr{K}$
of $n\times n$ and ($n>2)$ reciprocal matrix $M$ is equal to:

\begin{equation}
\mathscr{K}(M)=\underset{i,j,k\in\{1,\ldots,n\}}{\max}\left\{ \min\left\{ \left|1-\frac{m_{ij}}{m_{ik}m_{kj}}\right|,\left|1-\frac{m_{ik}m_{kj}}{m_{ij}}\right|\right\} \right\} \label{eq:1-koczkod_inc_idx}
\end{equation}

where $i,j,k=1,\ldots,n$ and $i\neq j\wedge j\neq k\wedge i\neq k$. 

There are also several different methods of deriving the weights vector
out of the matrix $M$ \cite{Saaty1998rbev,Ishizaka2006htdp}. Two
the most popular are the eigenvector method \cite{Saaty2008rmai}
and the geometric mean method. According to the first one, the output
$\mu$ (denoted as $\mu_{\text{EV}}$) is the rescaled principal eigenvector
of $M$, i.e.: 
\begin{equation}
\mu_{\textit{EV}}=\left[\frac{v_{1}}{s_{\textit{\textit{EV}}}},\ldots,\frac{v_{n}}{s_{\textit{EV}}}\right]^{T}\,\,\,\,\,\mbox{where}\,\,\,\, s_{\textit{EV}}=\underset{i=1}{\overset{n}{\sum}}v_{i}\label{eq:mu_EV_def}
\end{equation}

and $v=\left[v_{1},\ldots,v_{n}\right]^{T}$ is the principal eigenvector
of $M$. The second method \cite{Crawford1987tgmp} proposes the adoption
of rescaled geometric means of rows of $M$ as the output $\mu$ .
Thus, 
\begin{equation}
\mu_{\text{\textit{GM}}}=\left[\frac{g_{1}}{s_{\text{\textit{GM}}}},\ldots,\frac{g_{n}}{s_{\text{\textit{GM}}}}\right]\label{eq:mu_GV_def}
\end{equation}

where 
\begin{equation}
\, g_{i}=\left(\prod_{j=1}^{n}m_{ij}\right)^{\nicefrac{1}{n}}\,\,\,\,\,\mbox{and}\,\,\,\,\, s_{\text{\textit{\text{GM}}}}=\underset{i=1}{\overset{n}{\sum}}g_{i}\label{eq:mu_GV_supp_def}
\end{equation}

\[
\]

Other\emph{ }the priority deriving methods in the \emph{AHP} approach
can be found in \cite{Ishizaka2006htdp,Ishizaka2011rotm,Yuen2010ahpp}.

\section{The HRE Algorithm Approach\label{sec:Original-HRE-Algorithm}}

\begin{figure}[h]
\begin{centering}
\centering 
\def\svgwidth{0.80\columnwidth} 
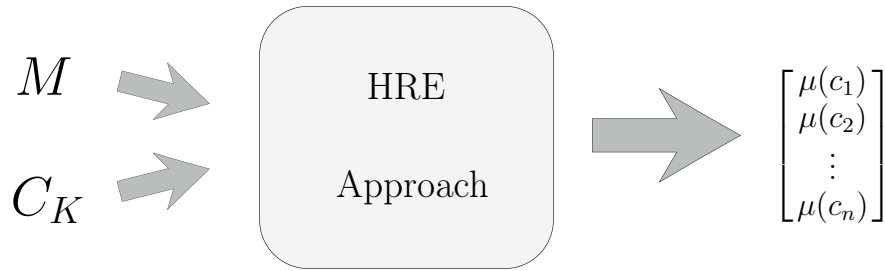
\par\end{centering}

\caption{\emph{HRE} approach input-output scheme. $C_{K}$ means the non-empty
set of reference concepts.}
\label{HRE_approach_fig}
\end{figure}

The \emph{HRE} approach to the rating estimation in the pairwise comparisons
method is based on a few intuitive heuristics. The first of them concerns
dividing the set of concepts into known (reference) and unknown elements.
Initially, $\mu$ is defined only for reference elements. Hence, only
these elements can be used to estimate $\mu$ for unknown elements.
With every subsequent step $\mu$ is specified for more and more elements.
Thus, increasing the number of elements could be taken into account
during calculations. The weights of initially known reference elements
remain unchanged. Thus, the subsequent updates affect only unknown
elements. In every step weights for unknown elements are determined
as the arithmetic mean of determined values and the appropriate ratios
(\ref{eq:2-main_hre_equation}). This iterative procedure forms an
averaging with respect to the reference heuristics (a more detailed
description in Sec. \ref{sub:Heuristics-of-averaging}). Therefore,
comparing with the eigenvalue based method, the \emph{HRE} approach
requires additional information about the reference elements (see
Fig. \ref{HRE_approach_fig}).

In fact, sometimes the \emph{HRE} procedure is equivalent to finding
a solution for some linear equation system. Hence, if this equation
system has an admissible solution, then its solution can be adopted
as the output of the \emph{HRE} algorithm. If not, the weight vector
needs to be determined with the help of the second heuristic (Sec.
\ref{sub:Heuristics-of-minimizing}) as explained later in the work.

In general, the pairwise comparisons method assumes that the input
matrix is reciprocal. It means that the ratio $m_{ij}$ expressing
the relative importance of $c_{i}$ compared to $c_{j}$ should be
the inverse of $m_{ji}$. Unfortunately, this assumption may not always
be met \cite{Koczkodaj2010odbi}. Since the situation when $m_{ij}\neq\nicefrac{1}{m_{ji}}$
is undesired, but possible in practice, the third heuristic proposes
a simple method for calculating the new values $\widehat{m}_{ij},\widehat{m}_{ji}$
so that they are mutually reciprocal and possibly close to the original
$m_{ij},m_{ji}$. The operation is called reciprocity restoration
and is applied to any matrix $M$, which is processed by the \emph{HRE}
algorithm. 

The fourth heuristic addresses the problem of incomplete data, where
not all the ratios $m_{ij}$ are known. Some of them can be recovered
based on the assumed reciprocity. However, if both $m_{ij}$ and its
counterpart $m_{ji}$ are unknown the reciprocity property does not
help. In such a case, either the missing ratios are reconstructed
\cite{Bozoki2010ooco,Koczkodaj1999mnei} so that the standard methods
can deal with the reconstructed matrix, or the procedure alone has
to deal with the problem of missing matrix entries. The iterative
\emph{HRE} approach does not need the matrix reconstruction. During
the course of the iterative procedure, the new ratio values are computed
using only those defined concepts that are reachable due to the availability
of an appropriate ratio. In other words, if some ratio is missing
all the multiplications which use the missing ratio are excluded from
the basic update formula (\ref{eq:2-main_hre_equation}).

\subsection{Heuristics of averaging with respect to reference values\label{sub:Heuristics-of-averaging}}

The iterative averaging approach presented in \cite{Kulakowski2013ahre}
assumes that the set of concepts $C=C_{K}\cup C_{U}$ and $C_{K}\cap C_{U}=\emptyset$,
where $C_{K}$ denotes concepts for which the actual value $\mu$
is initially known, and $C_{U}$ contains concepts for which the value
$\mu$ needs to be determined. The relation between different concepts
in $C$ is represented by $M$ so that in the case of the fully consistent
matrix it holds that $\mu(c_{i})m_{ji}=\mu(c_{j})$. Hence, for a
known, complete matrix $M$ and $c_{i}\in C_{K},c_{j}\in C_{U}$ determining
$c_{j}$ boils down to the performance of a single multiplication.
Since $M$ is usually inconsistent, the \emph{HRE} algorithm considers
$m_{ji}\mu(c_{i})$ as a sample of $\mu(c_{j})$, where the expected
value of $\mu(c_{j})$ is the arithmetic mean of the values $m_{ji}\mu(c_{i})$.
Of course, not all the values $\mu(c_{i})$ are defined at the very
beginning, but only those for which $c_{i}\in C_{K}$. Hence, in the
first step of the \emph{HRE} procedure, the values $\mu(c_{j})$ are
estimated only on the basis of the initially known concepts. However,
in the second step (assuming that $M$ is complete) all the other
values $\mu(c_{i})$ computed during the first step (for $i\neq j$
and $c_{i}\in C_{K}\cup C_{U}$) can be used to determine $\mu(c_{j})$.
Thus, for every concept $c_{j}\in C_{K}$ the r'th subsequent estimation
of $\mu_{r}(c_{j})$ computed by the \emph{HRE} iterative procedure
(see \cite{Kulakowski2013ahre}) meets the equation: 
\begin{equation}
\mu_{r}(c_{j})=\frac{1}{\left|C_{j}^{r-1}\right|}\overset{}{\underset{c_{i}\in C_{j}^{r-1}}{\sum}m_{ji}\mu_{r-1}(c_{i})}\label{eq:2-main_hre_equation}
\end{equation}

where 
\begin{equation}
C_{j}^{r-1}=\{c\in C:\mu_{r-1}(c)\,\,\,\mbox{is known and}\,\,\, c\neq c_{j}\},\,\,\,\mbox{and}\,\, C_{j}^{0}\overset{df}{=}C_{K}\label{eq:3-known_set_def}
\end{equation}

and $\left|C_{j}^{r-1}\right|$ is the cardinality (number of elements)
of $C_{j}^{r-1}$ \cite{Kulakowski2013ahre}. For simplicity, let
us assume that $C_{U}=\{c_{1},\ldots,c_{k}\}$ and $C_{K}=\{c_{k+1},\ldots,c_{n}\}$.
It turns out that the iterative procedure proposed in \cite{Kulakowski2013ahre}
follows the Jacobi iterative method for solving a linear equation
system in the form%
\footnote{The form of the linear equation system (\ref{eq:main_matrix_equation})
is more thoroughly explained in Appendix \ref{sec:Heuristics-of-averaging-forming-the-lin-eq-system}.%
}: 

\begin{equation}
A\mu=b\label{eq:main_matrix_equation}
\end{equation}

where the matrix $A$ is given as:

\begin{equation}
A=\left[\begin{array}{cccc}
1 & -\frac{1}{n-1}m_{1,2} & \cdots & -\frac{1}{n-1}m_{1,k}\\
-\frac{1}{n-1}m_{2,1} & 1 & \cdots & -\frac{1}{n-1}m_{2,k}\\
\vdots & \vdots & \vdots & \vdots\\
-\frac{1}{n-1}m_{k-1,1} & \cdots & \ddots & -\frac{1}{n-1}m_{k-1,k}\\
-\frac{1}{n-1}m_{k,1} & \cdots & -\frac{1}{n-1}m_{k,k-1} & 1
\end{array}\right]\label{eq:25-eq-1}
\end{equation}

vector of constant terms is 
\begin{equation}
b=\left[\begin{array}{c}
\frac{1}{n-1}m_{1,k+1}\mu(c_{k+1})+\ldots+\frac{1}{n-1}m_{1,n}\mu(c_{n})\\
\frac{1}{n-1}m_{2,k+1}\mu(c_{k+1})+\ldots+\frac{1}{n-1}m_{2,n}\mu(c_{n})\\
\vdots\\
\frac{1}{n-1}m_{k,k+1}\mu(c_{k+1})+\ldots+\frac{1}{n-1}m_{k,n}\mu(c_{n})
\end{array}\right]\label{eq:constant_terms_vector}
\end{equation}

and values that need to be determined are denoted as:

\begin{equation}
\mu^{T}=\left[\mu(c_{1}),\ldots,\mu(c_{k})\right]\label{eq:unknown_terms_vector}
\end{equation}

 The iteration matrix of the Jacobi method is given by:

\begin{equation}
B_{J}=D^{-1}(E+F)=I-D^{-1}A\label{eq:Jacobi_matrix_decomposition}
\end{equation}

The matrix $D$ is the diagonal matrix of the diagonal entries of
$A$, hence $D=D^{-1}=I$, whilst $E$ is the lower triangular matrix
of entries $e_{ij}=-\left(-\frac{1}{n-1}m_{i,j}\right)=-a_{ij}$,
and $F$ is the upper triangular matrix of entries $f_{ij}=-\left(-\frac{1}{n-1}m_{i,j}\right)=-a_{ij}$.
Therefore, the update equation (\ref{eq:2-main_hre_equation}) can
be written in the form:
\begin{equation}
\mu_{r}(c_{i})=\frac{1}{a_{ii}}\left[b_{i}-\sum_{j=1,j\neq i}^{k}a_{ij}\mu_{r-1}(c_{j})\right]=b_{i}+\sum_{j=1,j\neq i}^{k}\frac{1}{n-1}m_{ij}\mu_{r-1}(c_{j})\label{eq:jacobi_update_equation}
\end{equation}

When the matrix $A$ is strictly diagonally dominant by rows i.e.
$\left|a_{ii}\right|>\Sigma_{j=1}^{n}\left|a_{ij}\right|$ for $i\neq j$
and $i=1,\ldots,k$ then the Jacobi method is convergent \cite{Quarteroni2000nm}%
\footnote{Note that the Jacobi method is convergent also for $A$ strictly dominant
by columns \cite{Bagnara95aunified}%
}. In our case $a_{ii}=1$ for $i=1,\ldots,k$, hence the \emph{HRE}
procedure is convergent if 
\begin{equation}
1>\sum_{j=1,j\neq i}^{k}\left|a_{ij}\right|\label{eq:convergence_condition}
\end{equation}

for all $i=1,\ldots,k$. Bearing in mind that $a_{ij}=-\frac{1}{n-1}m_{ij}$
let us note the \emph{HRE} algorithm has a high chance to be convergent~if
the set $C_{U}$ is relatively small ($C_{K}$ is relatively large)
and $m_{ij}$ are not too large i.e. estimated values $\mu(c_{j})$
for $j=1,\ldots,k$ are similar. Both of these conditions are intuitive
and, in practice, are likely to be satisfied. The first of them reflects
the natural desire to provide the experts with rather more than the
lower number of known reference concepts. The second corresponds to
the common-sense observation that all the considered concepts should
be similar to each other, because then it is easy to compare them. 

The equation (\ref{eq:main_matrix_equation}) could also be solved
using direct methods. In such a case it has exactly one solution,
if the determinant of $A$ differs from $0$, i.e.:
\begin{equation}
\det(A)\neq0\label{eq:26-eq-det-1}
\end{equation}

Unfortunately, it may turn out that this unique solution $\mu=(\mu(c_{1}),\ldots,\mu(c_{k}))$
is not in $\mathbb{R}_{+}^{k}$. For instance, some values $\mu(c_{i})$
may be less than or equal $0$. In such a case the iterative approach
is not convergent (assuming that $\mu(c_{j})$ for $c_{j}\in C_{K}$
are strictly positive, the values $\mu(c_{i})$ for $i=1,\ldots,k$
must also be strictly positive%
\footnote{Note that all the components of the right side of \ref{eq:jacobi_update_equation}
are strictly positive.%
}). In such a case $\mu$ that meets (\ref{eq:main_matrix_equation})
cannot be adopted as the \emph{HRE} procedure output. Instead, the
\emph{HRE} procedure needs to be iterated a predetermined number of
times and the result $\mu$ needs to be chosen following the minimizing
estimation error heuristics (Sec. \ref{sub:Heuristics-of-minimizing}).
In the presented approach, only those $m_{ji}$ are determined by
experts for which at least one of the two $c_{i},c_{j}$ comes from
$C_{U}$. For two initially known concepts the value $m_{ji}$ is
just defined as $m_{ji}=\nicefrac{\mu(c_{j})}{\mu(c_{i})}$. Hence,
the matrix $M$ is always consistent in the part relating to the known
concepts i.e. $\forall c_{i},c_{j}\in C_{K}:m_{ji}\mu(c_{i})=\mu(c_{j})$. 

When the $\mu$ values are not initially known for any $c\in C$,
i.e. $C_{K}=\emptyset$, then for an arbitrarily selected $c_{i}$
the value $\mu(c_{i})$ might be set by the experimenter to $1$.
In such a case the \emph{HRE} procedure computes the relative order
$\mu$ of concepts from $C$ assuming that the weight $\mu(c_{i})$
is a unit. Since $c_{i}$ is treated as the reference element it must
be selected with special care. Their relationship with other concepts
have a reference meaning, hence they should be highly reliable and
well documented. 

The final weight vector $\mu_{\textit{\textit{HRE}}}$ is synthesized
by using $k$ values determined by solving (\ref{eq:main_matrix_equation})
and $n-k$ initially known reference values of concepts from $C_{K}$.
\begin{equation}
\mu_{\textit{HRE}}=\left[\mu(c_{1}),\ldots\mu(c_{k}),\mu(c_{k+1}),\ldots,\mu(c_{n})\right]^{T}\label{eq:mu_hre_def}
\end{equation}

Thus the rescaled form of $\mu_{\textit{HRE}}$ is:

\begin{equation}
\mu_{\textit{HREn}}=\left[\frac{\mu(c_{1})}{s_{\textit{HRE}}},\ldots,\frac{\mu(c_{n})}{s_{\textit{HRE}}}\right]\,\,\,\,\,\mbox{where}\,\,\,\,\, s_{\textit{HRE}}=\underset{i=1}{\overset{n}{\sum}}\mu(c_{i})\label{eq:mu_HRE_norm}
\end{equation}

\subsection{Heuristics of minimizing estimation error\label{sub:Heuristics-of-minimizing}}

The minimizing estimation error heuristics is proposed to deal with
the case when it is impossible to uniquely determine $\mu(c_{i})$
as the mean of $m_{ij}\mu(c_{j})$ for $i\neq j$ (the vector $\mu$
cannot be determined by solving%
\footnote{For the purpose of the \emph{HRE} approach only $\mu\in\mathbb{R}_{+}^{k}$
are admissible. %
} (\ref{eq:main_matrix_equation})). In tests, it was noticed that
the more often it happens, the higher the inconsistency. In such a
case, rather than solving (\ref{eq:main_matrix_equation}) someone
may try to find $\mu$ that minimizes the average absolute estimation
error, given as follows:

\begin{equation}
\widehat{e}_{\mu}=\frac{1}{\left|C_{U}\right|}\overset{}{\underset{c\in C_{U}}{\sum}e_{\mu}(c)}\label{eq:app_eq_1-1}
\end{equation}

where

\begin{equation}
e_{\mu}(c_{j})=\frac{1}{\left|C_{j}^{r-1}\right|}\overset{}{\underset{c_{i}\in C_{j}^{r-1}}{\sum}}\left|\mu(c_{j})-\mu(c_{i})\cdot m_{ji}\right|\label{eq:app_eq_2-1}
\end{equation}

The problem of minimizing $\widehat{e}_{\mu}$ is discussed in (Appendix
\ref{sec:minimizing-error-heuristics-explanation}). The preliminary
\emph{Monte Carlo} tests  show that for the relatively small inconsistency
(small $\mathscr{K}$) both: the minimizing estimation error heuristic
(as defined above) and the averaging with respect to the reference
values heuristic (Sec. \ref{sub:Heuristics-of-averaging}) lead to
very similar vectors $\mu$. When the inconsistency index $\mathscr{K}$
rises then the solutions provided by these two heuristics become increasingly
different. In general, it seems that the heuristic of averaging with
respect to the reference values is more useful in practice. However,
when the equation (\ref{eq:main_matrix_equation}) does not have an
admissible solution and there is an admissible $\mu$ minimizing (\ref{eq:app_eq_1-1}),
then the minimizing estimation error heuristic may be worth considering.

Certainly the search for the smallest $\widehat{e}_{\mu}$ makes sense
if both: solving (\ref{eq:main_matrix_equation}) and finding $\mu$,
which minimizes (\ref{eq:app_eq_1-1}) fail. Then the intermediate
\emph{HRE} iteration result with the minimal absolute estimation error
$\widehat{e}_{\mu_{r}}$ needs to be adopted as the output $\mu_{\textit{out}}$
of the \emph{HRE} procedure: 
\begin{equation}
\mu_{\textit{out}}=\left\{ \mu_{q}:\widehat{e}_{\mu_{q}}=\min\{\widehat{e}_{\mu_{1}},\ldots,\widehat{e}_{\mu_{r}}\}\right\} \label{eq:mu_out}
\end{equation}

Although $r$ - the total number of iterations has to be arbitrarily
set by an experimenter, in practice, it should be small enough (even
one or two iterations may be useful).

\subsection{Heuristics of reciprocity restoration}

According to these heuristics, the input \emph{PC} matrix $M$ should
be reciprocal to be processed by the \emph{HRE} procedure. Hence,
it should hold that $m_{ij}=\frac{1}{m_{ji}}$ for every two ratios
$m_{ij}$ and $m_{ji}$ in $M$. Therefore, if the matrix $M$ is
not reciprocal, it should be transformed to a similar but reciprocal
matrix. Let $\widehat{M}=\left[\widehat{m}_{ij}\right]$ be the new
\emph{PC} matrix obtained from $M=\left[m_{ij}\right]$ by replacing
entry $m_{ij}$ in $M$ by the geometric mean of this entry and its
(possibly reciprocal) counterpart i.e. $\widehat{m}_{ij}=\left(m_{ij}\frac{1}{m_{ji}}\right)^{\nicefrac{1}{2}}$.
It is easy to check that the new matrix $\widehat{M}$ is reciprocal.
Moreover, if $M$ is initially reciprocal then $\widehat{M}=M$. Therefore,
every \emph{PC} matrix $M$ calculated by the \emph{HRE} procedure
should be preprocessed in order to restore the lost reciprocity property.
If $M$ is reciprocal the preprocessed matrix should be identical
to $M$. If not, it is recommended to transform $M$ into $\widehat{M}$
according to the definition given above. A similar approach to the
lack of the reciprocity property has been discussed, for example,
in \cite{Fedrizzi2010otpv}. The geometric mean properties have been
discussed in \cite{Crawford1987tgmp}.

\subsection{Heuristics of missing data}

Sometimes there may be a situation that not all indispensable ratios
are defined. Then the resulting pairwise comparisons matrix $M$ is
incomplete and contains unknown values. In such a case the update
equation (\ref{eq:2-main_hre_equation}) cannot include the products
$m_{ji}\mu(c_{i})$ where $m_{ji}$ is not specified. Let us denote
$m_{ji}=$ ? if $m_{ji}$ is unspecified.

To handle this situation, the set of elements for which the $\mu_{r-1}$
values were known needs to be changed as follows:

\begin{equation}
C_{j}^{r-1}=\{c\in C:\mu_{r-1}(c)\,\,\,\mbox{is known,}\,\, c\neq c_{j}\,\,\,\mbox{and}\,\,\, c\neq c_{i}\,\,\,\mbox{when }m_{ji}=\mbox{?}\}\label{eq:3-known_set_def-1}
\end{equation}

Although incomplete, $M$ should be reciprocal. Hence the reciprocity
restoration procedure needs to be extended to the case when some ratios
are unknown. Thus, let us define: 
\begin{equation}
\widehat{m}_{ij}=\begin{cases}
\left(m_{ij}\frac{1}{m_{ji}}\right)^{\nicefrac{1}{2}} & \mbox{where}\,\,\, m_{ji}\,\,\mbox{and}\,\, m_{ij}\,\,\,\mbox{are specified in }M\\
m_{ij} & \mbox{where}\,\,\, m_{ji}\,\,\mbox{is unspecified in}\,\, M\\
\frac{1}{m_{ji}} & \mbox{where}\,\,\, m_{ij}\,\,\mbox{is unspecified in}\,\, M\\
? & \mbox{where}\,\,\, m_{ji}\,\,\mbox{and}\,\, m_{ij}\,\,\,\mbox{are unspecified in }M
\end{cases}\label{eq:reciprocity_decision_function}
\end{equation}

The \emph{HRE} algorithm equipped with the heuristics of missing data
can handle matrices to which other methods might not be applicable%
\footnote{Various methods have their extensions to enable them to handle such
cases, for example, the LSM extension can be found in \cite{Bozoki2008sotl}. %
}. The only limitation is the reachability of the unknown concepts
understood as the condition that for each unknown concept $c_{j}\in C_{U}$
there must exist at least one concept $c_{i}\in C_{K}$ with known
weight and a sequence of indices $i_{1},i_{2},\ldots,i_{q}$ such
that $m_{ii_{1}}\neq?,m_{i_{1}i_{2}}\neq?,\ldots,m_{i_{q}j}\neq?$,
where $i_{1},i_{2},\ldots,i_{q}\in\{1,2,\ldots,n\}$. Therefore,
the \emph{HRE} procedure is able to propose the value $\mu_{q}(c_{j})$
for $c_{j}\in C_{U}$ only if there is at least one $c_{r}\in C_{U}$
for which the product $m_{j,r}\mu_{q-1}(c_{r})$ is known. In the
case of an incomplete matrix the weights cannot be obtained by solving
a linear equation system as shown in (\ref{eq:main_matrix_equation}).
In particular, due to the missing data, the set of known values $C_{j}^{r-1}$
for $j$ such that $c_{j}\in C_{U}$ may change for the second and
subsequent iteration. Thus, the incomplete data requires an iterative
approach when every subsequent value of $\mu(c_{j})$ is estimated
according to the update rule (\ref{eq:2-main_hre_equation}). If the
procedure converges, a sufficiently accurate approximation might be
adopted as the output. If not, the one with the smallest $\widehat{e}_{\mu}$
from the several initial iteration results needs to be adopted as
the result of the procedure.

The missing data heuristics might be especially useful when a large
number of different concepts should be compared with each other. In
such a case the completion of all the ratios in the matrix $M$ might
be difficult, which may result in its incompleteness.

\section{Numerical examples\label{sec:Numerical-examples}}

Despite the fact that in the \emph{HRE} approach the priorities of
some concepts have to be initially known, the procedure might be used
(for caution) in any case. However, this will require the adoption
of arbitrarily selected elements as the reference concepts. The first
numerical example (from \cite{BanaeCosta2008acao}) demonstrates the
case when the standard \emph{PC} matrix is processed by the \emph{HRE}
algorithm and the arbitrary concept is chosen as the reference one.
The second example shows a typical situation for \emph{HRE}. There
is a non-empty set $C_{K}$ of the reference concepts and the set
$C_{U}$ consists of unknown elements. The third example addresses
the problem of non-reciprocal matrices and demonstrates how the heuristic
of reciprocity restoration works in practice. The last, fourth, example
deals with an incomplete \emph{PC} matrix. It uses an iterative version
of the \emph{HRE} procedure to derive the weight vector from $M$.
It is designed to demonstrate how the iterative \emph{HRE} procedure
equipped with the missing data heuristic may support incomplete \emph{PC}
data sets.

\subsection*{Example 1 (Case of verbal judgements)}

Let $c_{1},\ldots,c_{5}$ be a set of concepts for which the following
judgements were formulated by a person $J$: $c_{1}$ equally to moderately
dominates $c_{2}$, $c_{1}$ moderately dominates $c_{3}$ , $c_{1}$
strongly dominates $c_{4}$, $c_{1}$ extremely dominates $c_{5}$,
$c_{2}$ equally to moderately dominates $c_{3}$, $c_{2}$ moderately
to strongly dominates $c_{4}$, $c_{2}$ extremely dominates $c_{5}$,
$c_{3}$ equally to moderately dominates $c_{4}$, $c_{3}$ very strongly
dominates $c_{5}$, $c_{4}$ very strongly dominates $c_{5}$. Then,
adopting the method of converting verbal judgements into numbers proposed
in \cite{Saaty2005taha} the following \emph{PC} matrix is obtained: 

\begin{equation}
\mathsf{M}=\left[\begin{array}{ccccc}
1 & 2 & 3 & 5 & 9\\
\frac{1}{2} & 1 & 2 & 4 & 9\\
\frac{1}{3} & \frac{1}{2} & 1 & 2 & 8\\
\frac{1}{5} & \frac{1}{4} & \frac{1}{2} & 1 & 7\\
\frac{1}{9} & \frac{1}{9} & \frac{1}{8} & \frac{1}{7} & 1
\end{array}\right]\label{eq:example_1_pc_matric_m}
\end{equation}

The rescaled eigenvector $\mu_{\textit{EV}}$ (see \ref{eq:mu_EV_def})
corresponding to the maximal eigenvalue of $M$ is: 

\begin{equation}
\mu_{\textit{EV}}=\left[\begin{array}{ccccc}
0.426 & 0.281 & 0.165 & 0.101 & 0.027\end{array}\right]^{T}\label{eq:ex_1_eigen_value_rank}
\end{equation}

The geometric mean based weight vector (see \ref{eq:mu_GV_def}) for
$M$ is: 

\begin{equation}
\mu_{\textit{GM}}=\left[\begin{array}{ccccc}
0.424 & 0.284 & 0.169 & 0.098 & 0.026\end{array}\right]^{T}\label{eq:ex_1_gm_value_rank}
\end{equation}

As reported in \cite{BanaeCosta2008acao} the ranking $\mu_{\textit{EV}}$
does not meet the \emph{Condition of Order Preservation} (herein abbreviated
as \emph{COP}, see Appendix \ref{sec:Condition-of-order-appendix}).
In particular, since the value $m_{1,4}=4$ is smaller than $m_{4,5}=7$,
\emph{COP} also requires that $\frac{\mu_{\textit{EV}}(c_{1})}{\mu_{\textit{EV}}(c_{4})}<\frac{\mu_{\textit{EV}}(c_{4})}{\mu_{\textit{EV}}(c_{5})}$.
It is easy to calculate that $\frac{\mu_{\textit{EV}}(c_{1})}{\mu_{\textit{EV}}(c_{4})}=4.218$
and $\frac{\mu_{\textit{EV}}(c_{4})}{\mu_{\textit{EV}}(c_{5})}=3.741$
which is in contradiction with the second \emph{COP} postulate (\ref{eq:8-eq:cop-quantitative-cond}).
It is easy to check that for $\mu_{\textit{GM}}$ \emph{COP }does
not hold either.\emph{ }The eigenvalue based inconsistency index is
low and equals $CI=0.057$. In contrast, Koczkodaj's distance based
inconsistency index is high%
\footnote{The work \cite{Koczkodaj2010odbi} suggests that an acceptable threshold
of inconsistency $\mathscr{K}(M)$, for most practical applications,
turns out to be $\nicefrac{1}{3}$.%
} and equals $\mathscr{K}(M)=0.743$.

To calculate the rank using the \emph{HRE} approach when none of the
concepts are initially known (i.e. $C_{K}=\emptyset$), it is necessary
to choose some $c\in C_{U}$ and assign an arbitrary weight to it.
Thus, based on our knowledge about the problem domain, let us assume
that $c_{1}$ is a reference element ($C_{K}=\{c_{1}\}$ and $C_{U}=C_{U}\backslash\{c_{1}\}$)
and set $\mu(c_{1})=1$. (It is easy to check that for a rescaled
form of a weight vector $\mu$ the exact value assigned to $\mu(c_{1})$
is not important). Then, after the first \emph{HRE} iteration, the
matrix $A$ and vector $b$ are determined%
\footnote{In practice, the matrix $A$ can be obtained from the matrix $M$
by removing the rows and columns corresponding to concepts from $C_{K}$,
and multiplying the remaining values (except diagonal) by $-\nicefrac{1}{(n-1)}$.
The removed rows and columns form the vector $b$ as shown in (\ref{eq:constant_terms_vector}).%
}, 

\begin{equation}
A=\left[\begin{array}{cccc}
1 & -\frac{1}{n-1}m_{2,3} & -\frac{1}{n-1}m_{2,4} & -\frac{1}{n-1}m_{2,5}\\
-\frac{1}{n-1}m_{3,2} & 1 & -\frac{1}{n-1}m_{3,4} & -\frac{1}{n-1}m_{3,5}\\
-\frac{1}{n-1}m_{4,2} & -\frac{1}{n-1}m_{4,3} & 1 & -\frac{1}{n-1}m_{4,5}\\
-\frac{1}{n-1}m_{5,2} & -\frac{1}{n-1}m_{5,3} & -\frac{1}{n-1}m_{5,4} & 1
\end{array}\right],\,\,\,\, b=\left[\begin{array}{c}
\frac{1}{n-1}m_{2,1}\mu(c_{1})\\
\frac{1}{n-1}m_{3,1}\mu(c_{1})\\
\frac{1}{n-1}m_{4,1}\mu(c_{1})\\
\frac{1}{n-1}m_{5,1}\mu(c_{1})
\end{array}\right]\label{eq:case_1_matrix_symb}
\end{equation}

so that the equation (\ref{eq:main_matrix_equation}) takes the form:

\begin{equation}
\left[\begin{array}{cccc}
1 & -0.5 & -1 & -2.25\\
-0.125 & 1 & -0.5 & -2\\
-0.062 & -0.125 & 1 & -1.75\\
-0.028 & -0.031 & -0.036 & 1
\end{array}\right]\left[\begin{array}{c}
\mu(c_{2})\\
\mu(c_{3})\\
\mu(c_{4})\\
\mu(c_{5})
\end{array}\right]=\left[\begin{array}{c}
0.125\\
0.083\\
0.05\\
0.028
\end{array}\right]\label{eq:case_1_matrix}
\end{equation}

(Note that $\left|C_{U}\right|=4$ implies that the dimensions of
matrix $A$ are $4\times4$). Since $\det(A)\neq0$ and $\mu(c_{i})>0$
for $i=2,\ldots,5$ then the rescaled vector $\mu_{\textit{HREn}}$
obtained by solving (\ref{eq:main_matrix_equation}) is adopted as
the output of the \emph{HRE} algorithm (an iterative procedure leads
to the same solution). 

\begin{equation}
\mu_{\textit{HREn}}=\left[\begin{array}{ccccc}
0.368 & 0.311 & 0.182 & 0.11 & 0.028\end{array}\right]^{T}\label{eq:case_1_HRE_result}
\end{equation}

By examining all the possible cases, it is easy to check that the
weight vector $\mu_{\textit{HREn}}$ satisfies \emph{COP}. It is noteworthy
that all the three vectors: $\mu_{\textit{EV}},\mu_{\textit{GM}}$
and $\mu_{\textit{HREn}}$ preserve the same order of elements and
they differ only in intensities of preferences. Since the value $\mu(c_{1})$
is chosen arbitrarily by an experimenter, the obtained result has
only an ordinal meaning. As both vectors $\mu_{\textit{HREn}}$ and
$\mu_{\textit{HRE}}$ carry the same (ordinal) information it is convenient
to consider the rescaled vector $\mu_{\textit{HREn}}$.

\subsection*{Example 2 (Case with reference concept values)}

The immediate inspiration for the second example is the scientific
units evaluation in Poland. The proposed ranking algorithm \cite{MinisterNiSW2012rmmi}
is based on the pairwise comparisons paradigm although it does not
follow the \emph{AHP} approach. The reference scientific units (as
defined therein) are used to determine the scientific categories,
and thereby funding levels.

Let $c_{1},\ldots,c_{5}$ represent the hypothetical scientific units,
where two of them $c_{2}$ and $c_{3}$ are the reference units for
which the values $c_{2},c_{3}\in C_{K}$ are initially known and equal
$\mu(c_{2})=5$ and $\mu(c_{3})=7$. The analysis of the scientific
units $c_{1},c_{4}$ and $c_{5}$ with respect to the criterion $\mu$
allows the formulation of the following pairwise comparisons matrix:

\begin{equation}
\mathsf{M}=\left[\begin{array}{ccccc}
1 & \frac{3}{5} & \frac{4}{7} & \frac{5}{8} & \frac{1}{2}\\
\frac{5}{3} & 1 & \frac{5}{7} & \frac{5}{2} & \frac{10}{3}\\
\frac{7}{4} & \frac{7}{5} & 1 & \frac{7}{2} & 4\\
\frac{8}{5} & \frac{2}{5} & \frac{2}{7} & 1 & \frac{4}{3}\\
2 & \frac{3}{10} & \frac{1}{4} & \frac{3}{4} & 1
\end{array}\right]\label{eq:example_2_pc_matrix_m}
\end{equation}

The rescaled eigenvector $\mu_{\textit{EV}}$ (see \ref{eq:mu_EV_def})
and the rescaled geometric mean based vector $\mu_{\textit{GM}}$
(see \ref{eq:mu_GV_def}) for $M$ are as follows:

\begin{equation}
\mu_{\textit{EV}}=\left[\begin{array}{ccccc}
0.12 & 0.275 & 0.356 & 0.131 & 0.118\end{array}\right]^{T}\label{eq:ex_2_eigen_value_rank}
\end{equation}

and 

\begin{equation}
\mu_{\textit{GM}}=\left[\begin{array}{ccccc}
0.113 & 0.28 & 0.359 & 0.133 & 0.114\end{array}\right]^{T}\label{eq:ex_2_gm_value_rank}
\end{equation}

The \emph{HRE} approach requires the solution of the linear equation
system for $A$ and $b$ as follows%
\footnote{note that $\left|C_{U}\right|=3$ implies that the dimensions of matrix
$A$ are $3\times3$%
}:

\begin{equation}
A=\left[\begin{array}{ccc}
1 & -\frac{1}{n-1}m_{1,4} & -\frac{1}{n-1}m_{1,5}\\
-\frac{1}{n-1}m_{4,1} & 1 & -\frac{1}{n-1}m_{4,5}\\
-\frac{1}{n-1}m_{5,1} & -\frac{1}{n-1}m_{5,4} & 1
\end{array}\right],\,\,\,\, b=\left[\begin{array}{c}
\frac{1}{n-1}m_{1,2}\mu(c_{2})+\frac{1}{n-1}m_{1,3}\mu(c_{3})\\
\frac{1}{n-1}m_{4,2}\mu(c_{2})+\frac{1}{n-1}m_{4,3}\mu(c_{3})\\
\frac{1}{n-1}m_{5,2}\mu(c_{2})+\frac{1}{n-1}m_{5,3}\mu(c_{3})
\end{array}\right]\label{eq:case_1_matrix_symb-1}
\end{equation}

hence, numerically: 

\begin{equation}
\left[\begin{array}{ccc}
1 & -0.156 & -0.125\\
-0.4 & 1 & -0.333\\
-0.5 & -0.187 & 1
\end{array}\right]\left[\begin{array}{c}
\mu(c_{1})\\
\mu(c_{4})\\
\mu(c_{5})
\end{array}\right]=\left[\begin{array}{c}
1.75\\
1.0\\
0.812
\end{array}\right]\label{eq:case_2_matrix}
\end{equation}

The not rescaled $\mu_{\textit{HRE}}$ weight vector is: 

\begin{equation}
\mu_{\textit{HRE}}=\left[\begin{array}{ccccc}
2.527 & 5.0 & 7.0 & 2.88 & 2.616\end{array}\right]^{T}\label{eq:case_2_hre_result_unnorm}
\end{equation}

and after rescaling: 

\begin{equation}
\mu_{\textit{HREn}}=\left[\begin{array}{ccccc}
0.126 & 0.249 & 0.349 & 0.144 & 0.13\end{array}\right]^{T}\label{eq:case_2_hre_result_norm}
\end{equation}

The inconsistency indices are $CI=0.07$ (AHP) and $\mathscr{K}(M)=0.781$
(Koczkodaj).

It is easy to observe that in this hypothetical case the eigenvalue
vector $\mu_{\textit{EV}}$ also violates \emph{COP}. That is because
the ratio $m_{1,5}=\frac{1}{2}<1$, whilst $\frac{\mu_{\textit{EV}}(c_{1})}{\mu_{\textit{EV}}(c_{5})}>1$.
The $\mu_{\textit{GM}}$ and $\mu_{\textit{HRE}}$ do not violate
the first \emph{COP} postulate (Appendix \ref{sec:Condition-of-order-appendix}).
However, all the vectors $\mu_{\textit{EV}},\mu_{\textit{GM}}$ and
$\mu_{\textit{HRE}}$ do not meet the second \emph{COP} postulate.

\subsection*{Example 3 (Case of not reciprocal matrix)}

The third example concerns a situation when the \emph{PC} matrix is
almost consistent but not reciprocal. Due to the lack of reciprocity,
the use of the eigenvalue method as well as the geometric means method
might be disputed (these methods are designed for reciprocal matrices
\cite{Ishizaka2006htdp}). Hence, the values $\mu_{\textit{EV}}$
and $\mu_{\textit{GM}}$ are computed just for testing the robustness
and sensitivity of both methods to the incorrect data. 

Let $c_{1},\ldots,c_{4}$ represent four candidates for the position
of a manager in some production company. As different examiners have
been involved in the recruitment process, one examiner rated $c_{4}$
social skills twice as high as $c_{1}$, whilst another examiner,
while comparing skills $c_{1}$ to $c_{4}$, ruled that both candidates
are exactly on the same level. Assuming that in all other cases the
recruitment committee has ruled that all other candidates present
the same level of social skills, the \emph{PC} matrix $M$ representing
the problem may appear as follows:

\begin{equation}
\mathsf{M}=\left[\begin{array}{cccc}
1 & 1 & 1 & 1\\
1 & 1 & 1 & 1\\
1 & 1 & 1 & 1\\
2 & 1 & 1 & 1
\end{array}\right]\label{eq:example_3_pc_matric_m}
\end{equation}

An attempt to calculate the eigenvector based or geometric mean based
rank leads to the following vectors: 

\begin{equation}
\mu_{EV}=\left[\begin{array}{cccc}
0.236 & 0.236 & 0.236 & 0.292\end{array}\right]^{T}\label{eq:ex_3_eigen_value_rank}
\end{equation}

\begin{equation}
\mu_{GM}=\left[\begin{array}{cccc}
0.239 & 0.239 & 0.239 & 0.284\end{array}\right]^{T}\label{eq:ex_3_GM_rank}
\end{equation}

For the purpose of the \emph{HRE} algorithm the reciprocity property
of $M$ must be restored. Thus, according to the heuristics of reciprocity
restoration $M$ is transformed to $\widehat{M}$ in the form:

\begin{equation}
\mathsf{\widehat{M}}=\left[\begin{array}{cccc}
1 & 1 & 1 & 0.707\\
1 & 1 & 1 & 1\\
1 & 1 & 1 & 1\\
1.414 & 1 & 1 & 1
\end{array}\right]\label{eq:example_3_matrix_after_r_restore}
\end{equation}

then $\widehat{M}$ is processed following procedures formulated in
(Sec. \ref{sub:Heuristics-of-averaging} and \ref{sub:Heuristics-of-minimizing}).
Since $C_{k}$ cannot be empty, then let us adopt $c_{1}$ as the
reference element i.e. $C_{K}=C_{K}\cup\{c_{1}\}$ and $\mu(c_{1})=1$.
Then, the matrix $A$ and vector $b$ can be determined, 

\begin{equation}
A=\left[\begin{array}{ccc}
1 & -\frac{1}{n-1}\widehat{m}_{2,3} & -\frac{1}{n-1}\widehat{m}_{2,4}\\
-\frac{1}{n-1}\widehat{m}_{3,2} & 1 & -\frac{1}{n-1}\widehat{m}_{3,4}\\
-\frac{1}{n-1}\widehat{m}_{4,2} & -\frac{1}{n-1}\widehat{m}_{4,3} & 1
\end{array}\right],\,\,\,\, b=\left[\begin{array}{c}
\frac{1}{n-1}\widehat{m}_{2,1}\mu(c_{1})\\
\frac{1}{n-1}\widehat{m}_{3,1}\mu(c_{1})\\
\frac{1}{n-1}\widehat{m}_{4,1}\mu(c_{1})
\end{array}\right]\label{eq:case_1_matrix_symb-2}
\end{equation}

thus, to determine the vector $\mu_{\textit{HRE}}$ the following
linear equation system needs to be solved:

\begin{equation}
\left[\begin{array}{ccc}
1 & -0.333 & -0.333\\
-0.333 & 1 & -0.333\\
-0.333 & -0.333 & 1
\end{array}\right]\left[\begin{array}{c}
\mu(c_{2})\\
\mu(c_{3})\\
\mu(c_{4})
\end{array}\right]=\left[\begin{array}{c}
0.333\\
0.333\\
0.471
\end{array}\right]\label{eq:example_3_linear_eq}
\end{equation}

The rescaled \emph{HRE} weight vector is:

\begin{equation}
\mu_{\textit{HREn}}=\left[\begin{array}{cccc}
0.227 & 0.25 & 0.25 & 0.273\end{array}\right]^{T}\label{eq:example_3_normalized_vector}
\end{equation}

Although all the tested methods rate the $c_{4}$ candidate higher
than the others, only the \emph{HRE} method rates $c_{1}$ below the
average. Hence, only the \emph{HRE} algorithm meets \emph{COP, }i.e.
$m_{4,1}>m_{4,2}\Rightarrow\frac{\mu(c_{4})}{\mu(c_{1})}>\frac{\mu(c_{4})}{\mu(c_{2})}$
is met only by $\mu_{\textit{HRE}}$ (let us note that $\mu_{\textit{EV}}(c_{1})=\mu_{\textit{EV}}(c_{2})$
$ $as well as $\mu_{\textit{GM}}(c_{1})=\mu_{\textit{GM}}(c_{2})$,
thus $\nicefrac{\mu_{\textit{EV}}(c_{4})}{\mu_{\textit{EV}}(c_{1})}=\nicefrac{\mu_{\textit{EV}}(c_{4})}{\mu_{\textit{EV}}(c_{2})}$
and $\nicefrac{\mu_{\textit{GM}}(c_{4})}{\mu_{\textit{GM}}(c_{1})}=\nicefrac{\mu_{\textit{GM}}(c_{4})}{\mu_{\textit{GM}}(c_{2})}$).
Although the eigenvalue and geometric mean methods have a problem
with \emph{COP} when $M$ is not reciprocal, it should be noted that
they have no problems with \emph{COP} for $\widehat{M}$. This may
suggest that the reciprocity restoration heuristic might be useful
also for other weight derivation methods.

\subsection*{Example 4 (Case of incomplete matrix)}

The fourth example represents situations where some data are missing.
The known ratios representing the relative importance of concepts
were placed into the matrix $M$. Question marks at the intersection
of row $i$ and column $j$ in the matrix (\ref{eq:example_4_pc_matric_m})
mean unknown values $m_{ij}$. The immediate inspiration for this
example was an observation of the meta analysis process in biochemistry
\cite{Kulakowska2011mama} where the number and diversity of analyzed
factors make drawing the final conclusions difficult or even impossible. 

Let us consider the four drugs $c_{1},\ldots,c_{4}$ with proven efficacy
in controlling the disease $X$. Based on the available scientific
articles, Dr H. came to the conclusion that $c_{1}$ and $c_{2}$
have similar efficacy, the same for $c_{3}$ and $c_{4}$. He also
came across research showing that in some cases $c_{2}$ is two times
more effective than $c_{3}$, and also $c_{4}$ fails three times
more likely than $c_{1}$. Unfortunately, Dr. H. found no studies
comparing the therapeutic effect of drugs in pairs $(c_{1},c_{3})$
and $(c_{2},c_{4})$. Therefore the \emph{PC} matrix $M$ (\ref{eq:example_4_pc_matric_m})
prepared by Dr H. looks like as follows:

\begin{equation}
\mathsf{M}=\left[\begin{array}{cccc}
1 & 1 & ? & ?\\
? & 1 & 2 & ?\\
? & ? & 1 & ?\\
\frac{1}{3} & ? & 1 & 1
\end{array}\right]\label{eq:example_4_pc_matric_m}
\end{equation}

The drug $c_{1}$ is very popular, so there are many studies on its
efficacy. There are also some studies that compare efficacy of $c_{1}$
and $c_{2}$ but $c_{2}$ is less popular. Since $c_{1}$ is the most
popular drug on $X$, and what follows, the relationship between $c_{1}$
and $c_{2}$ have been most extensively tested, then $c_{1}$ has
been adopted as the reference concept, i.e. $C_{K}=C_{K}\cup\{c_{1}\}$
and $\mu(c_{1})=1$. The \emph{HRE} procedure, applied to $M$ (a
reciprocity restoration included) converges to: 

\begin{equation}
\mu_{\textit{HREn}}=\left[\begin{array}{cccc}
0.369 & 0.338 & 0.154 & 0.138\end{array}\right]^{T}\label{eq:example_4_normalized_vector}
\end{equation}

Thus, the most recommended cure for X is $c_{1}$, then $c_{2},c_{3}$
and $c_{4}$. It should be noted that the proposed weights by the
\emph{HRE} algorithm are in line with \emph{COP}. For example, if
$m_{2,3}=2>1$ then also $\frac{\mu_{\textit{HRE}}(c_{1})}{\mu_{\textit{HRE}}(c_{3})}=2,396>1$,
Similarly, $m_{4,1}=\frac{1}{3}<1$ then also $\frac{\mu_{\textit{HRE}}(c_{4})}{\mu_{\textit{HRE}}(c_{1})}=0,374<1$
etc. Due to the incompatible input matrix format, the eigenvalue method
and the geometric mean method could not be used in this case%
\footnote{There are several approaches that address the problem of incomplete
PC data. See e.g. \cite{Bozoki2010ooco,Fedrizzi2007ipca,Koczkodaj1999mnei}%
}.

\section{Summary\label{sec:Summary}}

The quality of the results achieved using the \emph{HRE} approach
is inextricably linked to input data quality. According to the popular
adage ``garbage in, garbage out'', when data are bad even the best
algorithm is not able to provide good output. In the case of heuristic
algorithms, the domain of applicability depends on the adopted heuristics.
Despite the promising results for different types (and different quality)
of input data, the application area of the \emph{HRE} approach has
only been sketched. It is therefore necessary to conduct further research
to better define assumed heuristics and the situations in which they
may be most useful. In particular, relationships between different
formulations of data inconsistency levels and the priority estimation
quality seem to be very interesting. 

The \emph{HRE} approach presented in the article is based on the iterative
\emph{HRE} algorithm primarily formulated in \cite{Kulakowski2013ahre}.
The heuristics indicated are much more thoroughly analyzed in this
work. In particular, the heuristic of averaging with respect to the
reference value and the heuristic of minimizing estimation error are
given in the general form as the linear equation system solving problems.
The new useful heuristic of reciprocity restoration has been introduced
and the incomplete \emph{PC} matrix problem has been addressed. The
presented theoretical considerations are accompanied by four numerical
examples demonstrating different situations in which the proposed
solution might be helpful. The \emph{HRE} approach tries to complement
other methods. It has been designed to help estimation of the relative
order of concepts when a non-empty reference subset of concepts is
known (or a set of such can be readily determined). Therefore, with
this new application area, it may be of interest to a wide range of
both researchers and practitioners.

\subsection*{Acknowledgements}

I would like to thank Dr Jaros\l{}aw W\k{a}s for reading the first
version of this work, his comments and corrections. I am also grateful
to Prof. Antoni Lig\k{e}za for valuable discussions and constant support
and encouragement. Special thanks are due to Ian Corkill for his
editorial help.

\bibliographystyle{plain}
\bibliography{papers_biblio_reviewed}

\begin{thebibliography}{10}

\bibitem{Bagnara95aunified}
R.~Bagnara.
\newblock A unified proof for the convergence of jacobi and gauss-seidel
  methods.
\newblock {\em SIAM Review}, 37, 1995.

\bibitem{BanaeCosta2008acao}
C.~A. Bana~e Costa and J.~Vansnick.
\newblock {A critical analysis of the eigenvalue method used to derive
  priorities in AHP}.
\newblock {\em European Journal of Operational Research}, 187(3):1422--1428,
  June 2008.

\bibitem{Barzilai1994arrn}
J.~Barzilai and B.~Golany.
\newblock {AHP rank reversal, normalization and aggregation rules}.
\newblock {\em INFOR - Information Systems and Operational Research},
  32(2):57--64, 1994.

\bibitem{Kulakowska2011mama}
A.~Bodzo{\'n}-Ku{\l}akowska, K.~Ku{\l}akowski, A.~Drabik, A.~Moszczynski,
  J.~Silberring, and P.~Suder.
\newblock {Morphinome--a meta-analysis applied to proteomics studies in
  morphine dependence.}
\newblock {\em Proteomics}, 11(1):5--21, January 2011.

\bibitem{Bozoki2008sotl}
S.~Boz{\'o}ki.
\newblock {Solution of the least squares method problem of pairwise comparison
  matrices}.
\newblock {\em Central European Journal of Operations Research},
  16(4):345--358, 2008.

\bibitem{Bozoki2010ooco}
S.~Boz{\'o}ki, J.~F{\"u}l{\"o}p, and L.~R{\'o}nyai.
\newblock On optimal completion of incomplete pairwise comparison matrices.
\newblock {\em Mathematical and Computer Modelling}, 52(1--2):318 -- 333, 2010.

\bibitem{Bozoki2008osak}
S.~Boz{\'o}ki and T.~Rapcsak.
\newblock {On Saaty's and Koczkodaj's inconsistencies of pairwise comparison
  matrices}.
\newblock {\em Journal of Global Optimization}, 42(2):157--175, 2008.

\bibitem{Brunelli2013iifp}
M.~Brunelli, L.~Canal, and M.~Fedrizzi.
\newblock {Inconsistency indices for pairwise comparison matrices: a numerical
  study}.
\newblock {\em Annals of Operations Research}, February 2013.

\bibitem{Colomer2011rlfa}
J.~M. Colomer.
\newblock {Ramon Llull: from `Ars electionis' to social choice theory}.
\newblock {\em Social Choice and Welfare}, 40(2):317--328, October 2011.

\bibitem{Crawford1987tgmp}
G.~B. Crawford.
\newblock The geometric mean procedure for estimating the scale of a judgement
  matrix.
\newblock {\em Mathematical Modelling}, 9(3--5):327 -- 334, 1987.

\bibitem{Dyer1990rota}
J.~S. Dyer.
\newblock {Remarks on the analytic hierarchy process}.
\newblock {\em Management Science}, 36(3):249--258, 1990.

\bibitem{Fechner1966eop}
G.~T. Fechner.
\newblock {\em Elements of psychophysics}, volume~1.
\newblock Holt, Rinehart and Winston, New York, 1966.

\bibitem{Fedrizzi2010otpv}
M.~Fedrizzi and M.~Brunelli.
\newblock On the priority vector associated with a reciprocal relation and a
  pairwise comparison matrix.
\newblock {\em Journal of Soft Computing}, 14(6):639--645, January 2010.

\bibitem{Fedrizzi2007ipca}
M.~Fedrizzi and S.~Giove.
\newblock {Incomplete pairwise comparison and consistency optimization}.
\newblock {\em European Journal of Operational Research}, 183(1):303--313,
  2007.

\bibitem{Greco2011fk}
S.~Greco, B.~Matarazzo, and R.~S{\l}owi{\'n}ski.
\newblock Dominance-based rough set approach on pairwise comparison tables to
  decision involving multiple decision makers.
\newblock In JingTao Yao, Sheela Ramanna, Guoyin Wang, and Zbigniew Suraj,
  editors, {\em Rough Sets and Knowledge Technology}, volume 6954 of {\em
  Lecture Notes in Computer Science}, pages 126--135. Springer Berlin
  Heidelberg, 2011.

\bibitem{Ho2008iahp}
William Ho.
\newblock {Integrated analytic hierarchy process and its applications - A
  literature review}.
\newblock {\em European Journal of Operational Research}, 186(1):18--18, March
  2008.

\bibitem{Ishizaka2009ahpa}
A.~Ishizaka and A.~Labib.
\newblock {Analytic hierarchy process and expert choice: Benefits and
  limitations}.
\newblock {\em OR Insight}, 22(4):201--220, 2009.

\bibitem{Ishizaka2011rotm}
A.~Ishizaka and A.~Labib.
\newblock {Review of the main developments in the analytic hierarchy process}.
\newblock {\em Expert Systems with Applications}, 38(11):14336--14345, October
  2011.

\bibitem{Ishizaka2006htdp}
A.~Ishizaka and M.~Lusti.
\newblock {How to derive priorities in AHP: a comparative study}.
\newblock {\em Central European Journal of Operations Research},
  14(4):387--400, December 2006.

\bibitem{Janicki2012oapc}
R~Janicki and Y.~Zhai.
\newblock On a pairwise comparison-based consistent non-numerical ranking.
\newblock {\em Logic Journal of the IGPL}, 20(4):667--676, 2012.

\bibitem{Koczkodaj1993ando}
W.~W. Koczkodaj.
\newblock A new definition of consistency of pairwise comparisons.
\newblock {\em Math. Comput. Model.}, 18(7):79--84, October 1993.

\bibitem{Koczkodaj1999mnei}
W.~W. Koczkodaj, M.~W. Herman, and M.~Orlowski.
\newblock {Managing Null Entries in Pairwise Comparisons}.
\newblock {\em Knowledge and Information Systems}, 1(1):119--125, 1999.

\bibitem{Koczkodaj1999caca}
W.~W. Koczkodaj and M.~Or{\l}owski.
\newblock {Computing a consistent approximation to a generalized pairwise
  comparisons matrix}.
\newblock {\em Computers {\&} Mathematics with Applications}, 37(3):79--85,
  1999.

\bibitem{Koczkodaj2010odbi}
W.~W. Koczkodaj and S.~J. Szarek.
\newblock {On distance-based inconsistency reduction algorithms for pairwise
  comparisons}.
\newblock {\em Logic Journal of the IGPL}, 18(6):859--869, October 2010.

\bibitem{Kulakowski2013ahre}
K.~Ku{\l}akowski.
\newblock A heuristic rating estimation algorithm for the pairwise comparisons
  method.
\newblock {\em Central European Journal of Operations Research}, pages 1--17,
  2013.

\bibitem{Liberatore2008tahp}
M.~J. Liberatore and R.~L. Nydick.
\newblock {The analytic hierarchy process in medical and health care decision
  making: A literature review}.
\newblock {\em European Journal of Operational Research}, 189(1):14--14, August
  2008.

\bibitem{Mikhailov2003dpff}
L.~Mikhailov.
\newblock {Deriving priorities from fuzzy pairwise comparison judgements}.
\newblock {\em Fuzzy Sets and Systems}, 134(3):365--385, March 2003.

\bibitem{MinisterNiSW2012rmmi}
{Ministry of Science and Higher Education}.
\newblock {Regulation on principles of science financing (Polish: Rozporz{\k
  a}dzenie Ministra Nauki i Szkolnictwa Wy{\.z}szego w sprawie kryteri{\'o}w i
  trybu przyznawania kategorii naukowej jednostkom naukowym)}.
\newblock {\em Dziennik Ustaw Rzeczypospolitej Polskiej}, 877, 2012.

\bibitem{Quarteroni2000nm}
A.~Quarteroni, R.~Sacco, and F.~Saleri.
\newblock {\em Numerical mathematics}.
\newblock Springer Verlag, 2000.

\bibitem{Saaty1977asmf}
T.~L. Saaty.
\newblock A scaling method for priorities in hierarchical structures.
\newblock {\em Journal of Mathematical Psychology}, 15(3):234 -- 281, 1977.

\bibitem{Saaty1990aeot}
T.~L. Saaty.
\newblock {An Exposition on the AHP in Reply to the Paper "Remarks on the
  Analytic Hierarchy Process"}.
\newblock {\em Management Science}, 36(3):259--268, March 1990.

\bibitem{Saaty2005taha}
T.~L. Saaty.
\newblock The analytic hierarchy and analytic network processes for the
  measurement of intangible criteria and for decision-making.
\newblock In {\em Multiple Criteria Decision Analysis: State of the Art
  Surveys}, volume~78 of {\em International Series in Operations Research and
  Management Science}, pages 345--405. Springer New York, 2005.

\bibitem{Saaty2008rmai}
T.~L. Saaty.
\newblock {Relative Measurement and Its Generalization in Decision Making. Why
  Pairwise Comparisons are Central in Mathematics for the Measurement of
  Intangible Factors. The Analytic Hierarchy/Network Process}.
\newblock {\em Estad{\'\i}stica e Investigaci{\'o}n Operativa / Statistics and
  Operations Research (RACSAM)}, 102:251--318, November 2008.

\bibitem{Saaty1998rbev}
T.~L. Saaty and G.~Hu.
\newblock {Ranking by eigenvector versus other methods in the analytic
  hierarchy process}.
\newblock {\em Applied Mathematics Letters}, 11(4):121--125, 1998.

\bibitem{Schlager2001sait2}
N.~Schlager and J.~Lauer, editors.
\newblock {\em Science and its times: understanding the social significance of
  scientific discovery}, volume~2.
\newblock Schlager Information Group, 2000.

\bibitem{Smith2004aada}
J.~E. Smith and D.~Von~Winterfeldt.
\newblock {Anniversary article: decision analysis in management science}.
\newblock {\em Management Science}, 50(5):561--574, 2004.

\bibitem{Subramanian2012aroa}
N.~Subramanian and R.~Ramanathan.
\newblock {A review of applications of Analytic Hierarchy Process in operations
  management}.
\newblock {\em International Journal of Production Economics}, 138(2):215--241,
  August 2012.

\bibitem{Thurstone27aloc}
L.~L. Thurstone.
\newblock A law of comparative judgment, reprint of an original work published
  in 1927.
\newblock {\em Psychological Review}, 101:266--270, 1994.

\bibitem{Vaidya2006ahpa}
O.~S. Vaidya and S.~Kumar.
\newblock {Analytic hierarchy process: An overview of applications}.
\newblock {\em European Journal of Operational Research}, 169(1):1--29,
  February 2006.

\bibitem{Yuen2010ahpp}
K.~K.~F. Yuen.
\newblock {Analytic hierarchy prioritization process in the AHP application
  development: A prioritization operator selection approach}.
\newblock {\em Appl. Soft Comput.}, 10(4):975--989, 2010.

\bibitem{Yuen2012mmpm}
K.~K.~F. Yuen.
\newblock {Membership Maximization Prioritization Methods for Fuzzy Analytic
  Hierarchy Process}.
\newblock {\em Fuzzy Optimization and Decision Making}, 11(2):113--133, June
  2012.

\bibitem{Yuen2013fcnp}
K.~K.~F. Yuen.
\newblock Fuzzy cognitive network process: Comparison with fuzzy analytic
  hierarchy process in new product development strategy.
\newblock {\em Fuzzy Systems, IEEE Transactions on}, PP(99):1--1, 2013.

\end{thebibliography}

\appendix

\section{About the heuristic of minimizing estimation error\label{sec:minimizing-error-heuristics-explanation}}

From the point of view of the heuristics of minimizing estimation
error, the best solution $\mu$ should minimize $\widehat{e}_{\mu}$

\begin{equation}
\widehat{e}_{\mu}=\frac{1}{\left|C_{U}\right|}\overset{}{\underset{c\in C_{U}}{\sum}e_{\mu}(c)}\label{eq:app_eq_1}
\end{equation}

where

\begin{equation}
e_{\mu}(c_{j})=\frac{1}{\left|C_{j}^{r-1}\right|}\overset{}{\underset{c_{i}\in C_{j}^{r-1}}{\sum}}\left|\mu(c_{j})-\mu(c_{i})\cdot m_{ji}\right|\label{eq:app_eq_2}
\end{equation}

The problem by replacing the absolute difference $\left|\mu(c_{j})-\mu(c_{i})\cdot m_{ji}\right|$
by the squared difference $(\mu(c_{j})-\mu(c_{i})\cdot m_{ji})^{2}$
leads to the equivalent one of finding the $\mu$ minimizing function
$f:\mathbb{R}_{+}^{k}\rightarrow\mathbb{R}$ given as: 

\begin{multline}
\,\,\,\,\,\,\,\,\,\,\,\,\,\,\,\,\,\,\,\,\,\,\,\,\,\,\,\,\,\,\,\,\,\,\,\,\,\,\,\,\,\,\,\overset{}{f(\mu(c_{1}),\ldots,\mu(c_{k}))=\underset{j\in I_{U}}{\sum}}\underset{i\in I_{U}\backslash\{j\}}{\sum}\left(\mu(c_{j})-\mu(c_{i})\cdot m_{ji}\right)^{2}\\
+\underset{j\in I_{U}}{\sum}\underset{i\in I_{K}}{\sum}\left(\mu(c_{j})-\mu(c_{i})\cdot m_{ji}\right)^{2}\,\,\,\,\,\,\,\,\,\,\,\,\,\,\,\,\,\,\,\,\,\,\,\,\,\,\,\,\,\,\,\,\,\,\,\,\label{eq:app_eq_eq_3}
\end{multline}

where $I_{U},I_{K}$ and $I_{C}$ denote the sets of indices of elements
from $C_{U},C_{K}$ and $C$ correspondingly%
\footnote{In particular it is assumed that $I_{U}=\{1,\ldots,k\}$%
}.

In order to determine the extremum of the function $f$, the following
linear equation system needs to be solved:
\begin{equation}
\left[\begin{array}{c}
\frac{\partial f}{\partial\mu(c_{1})}\\
\vdots\\
\frac{\partial f}{\partial\mu(c_{k})}
\end{array}\right]=0\label{eq:app_eq_4}
\end{equation}

where every single equation has the form:

\begin{multline}
\,\,\,\,\,\,\,\,\,\,\,\,\,\,\,\,\,\,\,\,\,\,\,\,\,\,\,\,\,\,\,\,\,\,\,\,\,\,\,\,\,\,\,\frac{\partial f}{\partial\mu(c_{j})}=2\cdot\left(\underset{i\in I_{U}\backslash\{j\}}{\sum}\left(\mu(c_{j})-\mu(c_{i})\cdot m_{ji}\right)-\right.\\
\left.-\underset{i\in I_{U}\backslash\{j\}}{\sum}\left(\mu(c_{i})-\mu(c_{j})\cdot m_{ij}\right)\cdot m_{ij}+\underset{i\in I_{K}}{\sum}\left(\mu(c_{j})-\mu(c_{i})\cdot m_{ji}\right)\right)=0\,\,\,\,\,\,\,\,\,\,\,\,\,\,\,\,\,\,\,\,\,\,\,\,\,\,\,\,\,\,\,\,\,\,\,\,\label{eq:app_eq_5}
\end{multline}

for $j\in I_{U}$. Hence, the above equation is equivalent to:
\begin{multline}
\left(n-1+\underset{i\in I_{U}\backslash\{j\}}{\sum}m_{ij}^{2}\right)\mu(c_{j})-\underset{i\in I_{U}\backslash\{j\}}{\sum}\left(m_{ji}+m_{ij}\right)\cdot\mu(c_{i})+\underset{i\in I_{K}}{\sum}\mu(c_{i})\cdot m_{ji}=0\\
\label{eq:app_eq_5-1}
\end{multline}

Dividing both sides of (\ref{eq:app_eq_5}) by $(n-1)$, and denoting
$\frac{1}{n-1}\underset{i\in I_{U}\backslash\{j\}}{\sum}m_{ij}^{2}\overset{df}{=}S_{j}$
it is easy to observe that (\ref{eq:app_eq_5-1}) turns into:

\begin{multline}
\,\,\,\,\,\,\,\,\,\,\,\,\,\,\,\,\,\,\,\,\,\,\,\,\,\,\,\,\,\,\,\,\,\,\,\,\,\,\,\,\,\,\,-\frac{m_{j1}+m_{1j}}{n-1}\mu(c_{1})-\frac{m_{j2}+m_{2j}}{n-1}\mu(c_{2})-\ldots\\
+\left(1+S_{j}\right)\mu(c_{j})-\frac{m_{jk}+m_{kj}}{n-1}\mu(c_{k})=b_{j}\,\,\,\,\,\,\,\,\,\,\,\,\,\,\,\,\,\,\,\,\,\,\,\,\,\,\,\,\,\,\,\,\,\,\,\label{eq:app_eq_6}
\end{multline}

Thus, finding the extremum point of $f$ boils down to solving the
following equation:

\begin{equation}
E\mu=b\label{eq:minimizin_error_problem_eq}
\end{equation}

where: 

\begin{equation}
E=\left[\begin{array}{cccc}
1+S_{1} & -\frac{m_{1,2}+m_{2,1}}{n-1} & \cdots & -\frac{m_{1,k}+m_{k,1}}{n-1}\\
-\frac{m_{2,1}+m_{1,2}}{n-1} & 1+S_{2} & \cdots & -\frac{m_{2,k}+m_{k,2}}{n-1}\\
\vdots & \vdots & \vdots & \vdots\\
\frac{m_{k-1,1}+m_{1,k-1}}{n-1} & \cdots & \ddots & -\frac{m_{k-1,k}+m_{k,k-1}}{n-1}\\
-\frac{m_{k,1}+m_{1,k}}{n-1} & \cdots & -\frac{m_{k,k-1}+m_{k-1,k}}{n-1} & 1+S_{k}
\end{array}\right]\label{eq:25-eq-1-1}
\end{equation}

and $b$ is defined as in (\ref{eq:constant_terms_vector}). It is
easy to show that the \emph{Hessian} matrix defined as: 
\begin{equation}
H_{ij}=\left[\frac{\partial^{2}f}{\partial\mu(c_{j})\partial\mu(c_{i})}\right]\label{eq:app_eq_7}
\end{equation}

equals:
\begin{equation}
H=2(n-1)E\label{eq:app_eq_8}
\end{equation}

Therefore, if $E$ is strictly diagonally dominant, then $H$ is also
strictly diagonally dominant. Since the diagonal entries of $H$ are
all positive, then $H$ is positively definite \cite[page 29]{Quarteroni2000nm}.
Thus, for $E$ strictly diagonally dominant the solution of (\ref{eq:minimizin_error_problem_eq})
is the minimum of $f$.

\section{Condition of order preservation\label{sec:Condition-of-order-appendix}}

Among the various criticisms raised at \emph{AHP} and the eigenvalue
method, a\emph{ Condition of Order Preservation} (\emph{COP}) postulate
\cite{BanaeCosta2008acao} seems to be one of the more interesting.
According to \emph{COP,} the output of the weight calculation method
should preserve the order as well as the intensity of preferences.
In other words, \emph{COP} is met by the weight vector $\mu$ if for
every four concepts $c_{1},\ldots,c_{4}\in C$ such that $c_{1}$
dominates $c_{2}$ more than $c_{3}$ dominates $c_{4}$ in $M$ i.e.
$m_{1,2}>1\wedge m_{3,4}>1\wedge m_{1,2}>m_{3,4}$, the following
two assertions are true:
\begin{enumerate}
\item Preservation of Order of Preference (\emph{POP})\\
\begin{equation}
\mu(c_{1})>\mu(c_{2})\label{eq:6-cop-qualitative-cond-1}
\end{equation}
\begin{equation}
\mu(c_{3})>\mu(c_{4})\label{eq:7-cop-qualitative-cond-2}
\end{equation}

\item Preservation of Order of Intensity of Preference (\emph{POIP})\\
\begin{equation}
\frac{\mu(c_{1})}{\mu(c_{2})}>\frac{\mu(c_{3})}{\mu(c_{4})}\label{eq:8-eq:cop-quantitative-cond}
\end{equation}

\end{enumerate}
COP does not depend on any concepts specific for the eigenvalue method.
It reflects the natural desire that the final ranking should be consistent
with the individual expert judgments. Thus, although \emph{COP} was
formulated with reference to the eigenvalue method, it might be used
as a quality test for any priority deriving methods, including the
\emph{HRE} approach.

\section{Heuristics of averaging with respect to reference values - the form of
the linear equation system\label{sec:Heuristics-of-averaging-forming-the-lin-eq-system}}

For simplicity, let us assume that $C_{U}=\{c_{1},\ldots,c_{k}\},\, C_{K}=\{c_{k+1},\ldots,c_{n}\}$.
The values $\mu$ for $c_{j}\in C_{K}$ are known, whilst the values
$\mu$ for elements of $C_{U}$ need to be estimated. The heuristics
of averaging with respect to the reference values assumes that for
every unknown $c_{j}\in C_{U}$ the value $\mu(c_{j})$ should
be estimated as the arithmetic mean of all the other values $\mu(c_{i})$
multiplied by factor $m_{ji}$: 
\begin{equation}
\mu(c_{j})=\frac{1}{n-1}\sum_{i=1,i\neq j}^{n}m_{ji}\mu(c_{i})\label{eq:append3_eq1}
\end{equation}

Thus, during the second and subsequent iterations the algorithm shown
in \cite{Kulakowski2013ahre} calculates the new estimation value
$\mu(c_{i})$ for each unknown concepts $c_{j}\in C_{U}$ according
to one of the following equations:

\begin{equation}
\begin{array}{ccc}
\mu(c_{1}) & = & \frac{1}{n-1}(m_{2,1}\mu(c_{2})+\dotfill+m_{n,1}\mu(c_{n}))\\
\mu(c_{2}) & = & \frac{1}{n-1}(m_{1,2}\mu(c_{1})+m_{3,2}\mu(c_{3})+\dotfill+m_{n,2}\mu(c_{n}))\\
\hdotsfor[1]{3}\\
\mu(c_{k}) & = & \frac{1}{n-1}\left(m_{1,k}\mu(c_{1})+\ldots+m_{k-1,k}\mu(c_{k-1})+m_{k+1,k}\mu(c_{k+1})+\ldots+m_{n,k}\mu(c_{n})\right)
\end{array}\label{eq:append3_eq2}
\end{equation}

Since the values $\mu(c_{k+1}),\ldots,\mu(c_{n})$ are known and constant
($c_{k+1},\ldots,c_{n}$ are the reference concepts), so they can
be grouped together. Let us denote:

\begin{equation}
b_{j}=\frac{1}{n-1}m_{k+1,j}\mu(c_{k+1})+\ldots+\frac{1}{n-1}m_{n,j}\mu(c_{n})\label{eq:append3_eq3}
\end{equation}

Thus, the linear equations system (\ref{eq:append3_eq2}) could be
written as: 

\begin{equation}
\begin{array}{ccc}
\mu(c_{1}) & = & \frac{1}{n-1}m_{2,1}\mu(c_{2})+\dotfill+\frac{1}{n-1}m_{k,1}\mu(c_{k})+b_{1}\\
\mu(c_{2}) & = & \frac{1}{n-1}m_{1,2}\mu(c_{1})+\frac{1}{n-1}m_{3,2}\mu(c_{3})+\ldots+\frac{1}{n-1}m_{k,2}\mu(c_{k})+b_{2}\\
\hdotsfor[1]{3}\\
\mu(c_{k}) & = & \frac{1}{n-1}m_{1,k}\mu(c_{1})+\dotfill+\frac{1}{n-1}m_{k-1,k}\mu(c_{k-1})+b_{k}
\end{array}\label{eq:append3_eq4}
\end{equation}

It is easy to see that the linear equation system (\ref{eq:append3_eq4})
forms the matrix equation (\ref{eq:main_matrix_equation}) where $A,b$
and $\mu$ are defined in (\ref{eq:25-eq-1}), (\ref{eq:constant_terms_vector})
and (\ref{eq:unknown_terms_vector}). Finding the solution of (\ref{eq:main_matrix_equation})
is equivalent to determine the values $\mu(c_{1}),\ldots,\mu(c_{k})$
with respect to the reference (known) concepts grouped in $C_{K}$.
\end{document}